\documentstyle[12pt]{article}
\begin{document}
\thispagestyle{empty}
\renewcommand{\thefootnote}{\fnsymbol{footnote}}
\newpage
\setcounter{page}{0}
\pagestyle{empty}
~\\
~
\vskip1.6cm
\begin{center}
{\LARGE {\bf{Two-dimensional $N=1,2$ Supersymmetric}
\\
~\vskip.3cm
{\bf Chiral and Dual Models}}}
\\
\vskip1cm
~\\
{\large{C.P. Constantinidis$^{(a1)}$, F.P. 
Devecchi$^{(b)}$ and F. Toppan$^{(a2)}$}}
~\\
\quad \\
\vskip.3cm
{\em{ ~$^{(a1,a2)}$ UFES, CCE Depto de F\'{\i}sica, Goiabeiras cep 29060-900,
Vit\'oria (ES) Brasil}}\\
~\\ 
{\em{ ~$^{(b)}$ UFPR, Depto de F\'{\i}sica, c.p. 19091, cep 81531-990,
Curitiba (PR) Brasil}}\\
\end{center}
\vskip.5cm
\centerline{\large {\bf Abstract}}
\vskip1cm
\noindent
Two-dimensional $N=1$,$2$ supersymmetric  
chiral models and their dual extensions are introduced
and canonically quantized.  Working within a superspace
formalism, the non-manifest invariance under $2D$-superPoincar\'e 
transformations is proven.
The $N=1,2$ superVirasoro algebras are recovered as current algebras. 
The non-anomalous quantum invariances
under $1D$-superdiffeomorphisms (for chiral models) and
$N=1,2$ superconformal transformations (for dual models)
are shown to be a consequence of an $N=1,2$ super-Coulomb
gas representation. 
\vskip.5cm
\vfill
\rightline{DF/UFES-P001/98}
{\em E-Mails:\\ 
~$^{(a1)}$  clisthen@cce.ufes.br\\
~$^{(b)}$ devecchi@fisica.ufpr.br\\
~$^{(a2)}$ toppan@cce.ufes.br}
\newpage
\pagestyle{plain}
\renewcommand{\thefootnote}{\arabic{footnote}}
\setcounter{footnote}{0}

\noindent{\section{Introduction.}}

\par

The Schwarz-Sen electromagnetic dual model \cite{Sen} 
is a four-dimensional member of a more general family of 
theories which are 
manifestly invariant under duality transformations
\cite{Hen1}. 
Its two-dimensional version is known as Tseytlin 
model \cite{Tsey}, introduced at first in the string theory
context. It is a conformal theory and as such can be 
decomposed into two independent (for left and 
right movers) Floreanini-Jackiw (FJ) \cite{Flo} chiral 
boson models.  FJ models have peculiar features which have
been extensively investigated in the last ten years \cite{Gir}. 
In particular, despite the fact they
are not-manifestly Lorentz-invariant, they turn out to
be $2$-dimensional Poincar\'e invariant. The 
quantum hamiltonian 
structure of the FJ model was analyzed in \cite{Gir},
while in \cite{CD} this analysis was extended to the 
Tseytlin model, proving in particular the closure of $2D$
Poincar\'e algebra both in the classical and in the 
quantum case. In \cite{CD} it was furthermore proposed a 
hamiltonian supersymmetric theory which coincides 
with a $2$-dimensional reduction of
the supersymmetric extension of the original 
Schwarz-Sen model \cite{Sen}. A complete analysis of its symmetries, 
as well as a manifest supersymmetric formulation, was 
however not carried out in that paper.
The purpose of our present work is to fully investigate
the properties of supersymmetric extensions of both
chiral (FJ) and dual (Tseytlin) models. Our aim is to 
provide the algebraic setting underlying 
dimensional reductions of supersymmetric $4$-dimensional dual
models. \par 
In this paper we construct for both FJ and Tseytlin 
models their $N=1$ and $N=2$ supersymmetric
extensions by using a superfield 
formalism\footnote{ In reality the ``$N=2$" Tseytlin model as 
a quantum mechanical system is globally $N=4$ supersymmetric. We 
discuss this point in detail in the following.}. We show
that their symmetries generate $N=1,2$ SuperVirasoro 
algebras and are in connection with the $N=1,2$ 
Coulomb gas formulation (see \cite{scg} and 
references therein). The closure of $2D$ ($N=1,2$) 
superPoincar\'e algebra is proven in both classical and quantum cases.\par
It is worth mentioning that the construction of  
supersymmetric extensions must be carefully performed,
which means their investigation is quite interesting. 
As an example we just mention that the equations
of motion for a system involving a chiral boson and a 
chiral fermion can be derived by using two
different hamiltonian pictures. Only one of 
them leads to a supersymmetric theory, 
while the other does not. Further topics of 
this kind will be discussed in the text.\par
The scheme of the paper is as follows:     
\par
In the next section we review the formulation of 
the bosonic FJ and Tseytlin models. Despite the fact that
most of the material presented is nowadays 
standard, some results presented are new.\par 
In section $3$ we introduce
the $N=1$ superfield formalism and derive the corresponding
super-FJ and superTseytlin models. Since it is not possible
to derive the equations of motion directly from a 
manifestly supersymmetric $2$-dimensional action, we supersymmetrize
the space coordinate only, leaving the time an ordinary
bosonic variable. Our formulation differs 
from a previously constructed version \cite{PST} in 
which one light-cone variable was supersymmetrized,
and is more suitable for analyzing the 
Tseytlin model which 
deals with both chiralities. 
The invariances under $1$-dimensional 
superdiffeomorphisms and $2$-dimensional
superPoincar\'e transformations are proven. The Dirac bracket
analysis is performed and the N\"other supercurrents are
derived. They realize an $N=1$ superVirasoro algebra
(for both chiralities in the Tseytlin model) with central
charge $c={\textstyle{3\over 2}}$ in the quantum case.\par 
The section $4$ deals with the $N=2$ extensions 
of the above models. They are 
constructed by making use of $N=2$ 
chiral and antichiral 
superfields\footnote{ The word is here 
employed to mean $N=2$ chirality and not 
the ordinary space-time chirality discussed above.}
and mimicking the procedure employed in the previous case. The total
field content (in the $N=2$ FJ model) 
consists of two ordinary chiral bosons and two ordinary chiral fermions. 
Invariances under $2$-dimensional $N=2$ superPoincar\'e
transformations and $1$-dimensional $N=2$ 
superdiffeomorphisms are proven. The Dirac's brackets for
the conserved N\"other currents generate an $N=2$ 
superVirasoro algebra with central charge $c=3$ in the quantum case. 

\vspace{0.2cm}
\noindent{\section{The bosonic FJ and Tseytlin models.}}

\par
Let us start introducing the bosonic FJ and Tseytlin models. 
They are defined in terms of the following lagrangian
densities:
\begin{eqnarray}
{\cal L}_{FJ} &=& \partial_0\phi\partial_1\phi - (\partial_1\phi)^2
\label{fjlagr} 
\end{eqnarray}
for the FJ model and
\begin{eqnarray}
{\cal L}_{Ts} &=& \partial_0\phi \partial_1{\tilde\phi}
+\partial_0{\tilde\phi}\partial_1\phi -(\partial_1\phi)^2
-(\partial_1{\tilde\phi})^2
\label{tslagr}
\end{eqnarray}
for the Tseytlin model.\par
The above lagrangians coincide with the ones given in the 
literature \cite{{Tsey},{Flo},{CD}} up to an 
overall normalization factor.\par
We work in the $2$-dimensional Minkowski spacetime; the time coordinate
$t$ will also be denoted as $x^0$ (${\partial_0}={\textstyle{\partial  
\over\partial t}}$) 
and the space coordinate $x$ as $x^1$ (${\partial_1}={\textstyle{
\partial\over\partial x}}$). We
will also make use of the light-cone coordinates $z_{\pm}$ defined as
\begin{eqnarray}
&& z_{\pm} = x \pm t,\quad\quad
\partial_{\pm} = {1\over 2}({\partial_1\pm\partial_0}).
\nonumber
\end{eqnarray}
The equations of motion are given by
\begin{eqnarray}
(\partial_0-\partial_1)\partial_1\phi &=& 0
\end{eqnarray}
in the FJ case and
\begin{eqnarray}   
&&{\partial_1}^2{\tilde\phi} -\partial_0\partial_1\phi =0
\, ,\quad\quad
{\partial_1}^2\phi -\partial_0\partial_1{\tilde \phi} =0
\end{eqnarray}
in the Tseytlin case.\\
The lagrangian density (\ref{tslagr}) is invariant under 
duality transformations, i.e. exchanging $\phi \leftrightarrow
{\tilde\phi}$. The Tseytlin model can be decomposed
into two (chiral and antichiral) FJ models as it is
evident from the positions
\begin{eqnarray}
\phi_\pm&=& {1\over \sqrt{2}} (\phi \pm {\tilde{\phi}})\, .
\end{eqnarray}
The lagrangian ${\cal L}_{Ts}$ can therefore be rewritten as
\begin{eqnarray}
{\cal L}_{Ts} &=& \partial_1\phi_+(\partial_1-\partial_0)\phi_+
+\partial_1\phi_-(\partial_1+\partial_0)\phi_-
\, .\end{eqnarray}
Both the FJ and the Tseytlin model are invariant under
Poincar\'e $2$-dimensional transformations, 
with $\phi$, ${\tilde\phi}$ transforming as scalar 
fields. A basic difference between the 
FJ theory and its dual version consists in 
the fact that in the chiral case the
hamiltonian $H\equiv P^0$ coincides with 
the space-translation generator $P^1$ while 
they are different for the Tseytlin lagrangian.\par
Another set of invariances is provided by a full class
of transformations dependent on a parameter $\lambda$.
Such kind of transformations are very well studied in the 
context of Coulomb gas picture \cite{DF}. However, they have not
been considered for the models we are dealing 
with. Indeed the  action correspondent to the 
FJ Lagrangian is invariant under the infinitesimal transformations:
\begin{eqnarray}
\delta_{\lambda} \phi(x,t) &=& 
\epsilon(z_+)\partial_1\phi + \lambda
\partial_1\epsilon (z_+)
\label{diff}
\end{eqnarray}
for any value of $\lambda$. The same transformation
applied to the $\phi_+$ field leaves 
invariant the Tseytlin action while a similar 
transformation applied on $\phi_-$ depends on an 
infinitesimal parameter ${\overline\epsilon} (z_-)$.\par
By expanding $\epsilon (z_+)$ in Laurent series
($\epsilon (z_+) = -\sum_{n\in{\bf Z}} \epsilon_n {z_+}^{n+1}$) we can
introduce for any $\lambda$
the operators $l_n(\lambda)$ given by
\begin{eqnarray}
l_n(\lambda) &=& -\left({z_+}^{n+1}\partial_1 + \lambda
(n+1) {z_+}^n\right)\, ,
\end{eqnarray}
therefore
\begin{eqnarray}
\delta_{\lambda}\phi &=& \sum_{n\in{\bf Z}} \epsilon_n
l_n(\lambda )\phi\, .
\end{eqnarray}
For any fixed value of $\lambda$ the algebra generated by the
$l_n(\lambda )$ operators is the $1$-dimensional diffeomorphisms 
algebra 
(centerless Virasoro algebra):
\begin{eqnarray}
\relax [l_n(\lambda) , l_m(\lambda) ] &=&
(n-m) l_{n+m} (\lambda)\, ,
\end{eqnarray}
therefore the set of 
invariances of the FJ model includes the
$1D$-diffeomorphisms, while in the Tseytlin case
we have the direct sum of two copies of $1D$-diffeomorphisms, one 
for each
chirality.\par
The analysis of the hamiltonian dynamics, the structure of primary
constraints and the construction of Dirac's brackets has been performed
in \cite{Gir} in the FJ case and in \cite{CD} in the Tseytlin case.\par
These results lead to the hamiltonian\footnote{ In the following formulae
the double dots denote the standard normal ordering; 
moreover for our purposes, in order to avoid complications arising
from boundary conditions, we assume the space coordinate $x$ being
compactified on a circle $S^1$ with periodic boundary conditions, or living
on
${\bf R}$ and the fields being fast-decreasing at the infinities.}  
\begin{eqnarray}
H_{FJ} &=& \int dx {\theta_{FJ}}^{00} (x)\, ,
\end{eqnarray}
with the current ${\theta_{FJ}}^{00}$ given by (in the quantum case)
\begin{eqnarray}
{\theta_{FJ}}^{00} &=& - : (\partial_1\phi )^2:
\quad .
\end{eqnarray}
The equal-time Dirac's brackets \cite{Dir}
are computed and give
\begin{eqnarray}
\relax [ \phi (x), \phi (y) ]_{D} &=&
{i\over 2}{\partial_y}^{-1}\delta (x-y)
\, ,
\end{eqnarray}
where the standard delta-function appears in the r.h.s.\par
In the Tseytlin case we have respectively
\begin{eqnarray}
H_{Ts} &=& -\int dx (:(\partial_1\phi_+)^2: + :(\partial_1\phi_-)^2:)\, ,
\end{eqnarray}
and the following Dirac's brackets
\begin{eqnarray}
&&\relax [ \phi_{\pm} (x), \phi_{\pm}(y)]_D=
\pm {i\over 2} {\partial_y}^{-1} \delta (x-y)\, ,
\quad\quad
\relax [\phi_+(x),\phi_-(y)]_D 
= 0\, .
\end{eqnarray}
The $2$-dimensional Poincar\'e algebra defined by the translation
generators $P^0$, $P^1$ and Lorentz boost $M$, with structure constants
\begin{eqnarray}
&&\relax [M,P^0] = i P^1\, ,\quad\quad
\relax [M,P^1] = i P^0
\label{poi}
\end{eqnarray}
(and vanishing otherwise), is reproduced by the conserved charges computed
with the standard N\"other methods. In the FJ case we
have
\begin{eqnarray}
&&P^1 = P^0 \equiv H_{FJ}\, ,\quad\quad
M = \int dx {\cal M} (x)\, ,
\end{eqnarray}  
with ${\cal M} (x)$ given by
\begin{eqnarray}
{\cal M} (x) &=& (x+t){\theta_{FJ}}^{00}(x,t)\, .
\end{eqnarray}
\par
We can compute the commutation 
relations of the ${\theta_{FJ}}^{00}(x)$ currents by using 
OPE techniques. The result is the following
\begin{eqnarray}
\relax [ {\theta_{FJ}}^{00} (x), {\theta_{FJ}}^{00} (y)] &=&
-{1\over 12}{\partial_y}^3\delta (x-y) +
2 i{\theta_{FJ}}^{00}(y)
\partial_y\delta (x-y) + i \partial_y{\theta_{FJ}}^{00}(y)\cdot
\delta (x-y)\, . \nonumber\\
&&
\label{virfj}  
\end{eqnarray}
The above algebra corresponds to the Virasoro algebra and the first term
in the r.h.s. gives the central extension. In the classical case such
term is not present and the algebra coincides 
with the algebra of $1$-dimensional diffeomorphisms. 
By reexpressing (\ref{virfj}) in standard quantum OPE form we
realize that the value of the central
charge $c$ corresponds to $c=1$.\par
By setting $P^0=P^1=P$ the $2D$-Poincar\'e algebra can be recovered from
the single commutation relation
\begin{eqnarray}
\relax [M,P] &=& i P\, .
\end{eqnarray}
We finish this part devoted to the FJ model by discussing its invariance
properties under $1D$-diffeomorphisms. For any given $\lambda$ 
the (\ref{diff}) transformations are generated by the N\"other currents
$\theta_\lambda (x,t)$ given by
\begin{eqnarray}
\theta_\lambda (x) &=& 
- :(\partial_1\phi)^2: + \lambda (\partial_1)^2 \phi 
\, .
\end{eqnarray}
The conserved charges are given by $L_n(\lambda)$,
\begin{eqnarray}
L_n(\lambda) &=& \int dx (x+t)^{(n+1)} \theta_\lambda (x,t)
\, .
\end{eqnarray}
The $L_n(\lambda)$ charges satisfy a closed algebra structure generated
by Dirac's brackets. It coincides with the Virasoro algebra with central
charge $c = 1-6i\lambda^2$: 
\begin{eqnarray}
\relax [L_n(\lambda), L_m(\lambda)]&=&
i (n-m) L_{n+m}(\lambda) -{c\over 12} n(n^2-1) \delta_{n+m,0}
\, .
\end{eqnarray}
The extra term $\lambda(\partial_1)^2\phi$ corresponds to 
the well-known Feigin-Fuchs term in the Coulomb gas picture \cite{DF}. 
Its purpose there consists 
in providing a bosonization for conformal theories with any specific value
of the central charge $c$. In the present context it tells us the following
feature of the model under consideration. It keeps an invariance under
$1D$-diffeomorphisms even in the quantum case because it is possible to 
find a particular value 
of $\lambda$ ($\lambda = \sqrt{{- i\over 6}}$) 
in such a way that the central charge is vanishing, 
which leads to a non-anomalous quantum
theory.\par
For what concerns the Tseytlin model its N\"other 
currents ${\theta_{Ts}}^{00}$, ${\theta_{Ts}}^{01}$ 
associated to the space-time translations can be decomposed as
\begin{eqnarray}
&&\theta_\pm = {\theta_{Ts}}^{00} \pm {\theta_{Ts}}^{01}=
\theta_\pm = -:(\partial_1\phi_\pm)^2:\quad ,
\end{eqnarray}
while the current ${\cal M}$ associated to the Lorentz boost is
\begin{eqnarray}
{\cal M} &=& x{\theta_{Ts}}^{00} + t {\theta_{Ts}}^{01}
\, .
\end{eqnarray}
The $2$-D Poincar\'e algebra (\ref{poi}) is realized
by the commutators of the conserved charges
\begin{eqnarray}
&& P^0 = \int dx {\theta_{Ts}}^{00},\quad\quad 
P^1= \int dx {\theta_{Ts}}^{01}, \quad\quad
M=\int dx {\cal M}
\, .
\end{eqnarray}
\par
The currents $\theta_\pm$ make explicit
the fact that the model under consideration is
conformally invariant since their commutation relations
satisfy the following algebraic relations
\begin{eqnarray}
\relax [\theta_\pm (x), \theta_\pm (y)]
&=& -{1/12} {\partial_y}^3\delta (x-y) + 2i\theta_\pm(y)
\partial_y\delta(x-y) + i\partial_y\theta_\pm (y)\cdot
\delta(x-y)\, ,\nonumber\\
\relax [\theta_+ (x), \theta_- (y)]&=& 0\, ,
\end{eqnarray}
which correspond to two separated copies (one for each chirality) of 
the Virasoro algebra, both with central charge $c=1$ in the quantum case. 
We wish to mention
that this analysis corrects a statement in \cite{CD} 
claiming the absence of the central extension 
(the proof there furnished of the Poincar\'e invariance 
remains valid because not affected by the presence
of the central term).

\vspace{0.2cm}
\noindent{\section{The $N=1$ supersymmetric FJ and dual models.}}

\par
The theory of a chiral boson $\phi$ and a chiral 
fermion $\psi $ consists in the system of equations of motion
\begin{eqnarray}
&&\partial_-\partial_1\phi = 0\, ,\quad\quad \partial_-\psi = 0\, .
\label{eqmo}
\end{eqnarray}
The above system of equations can be recovered from a 
single superfield equation where both spacetime 
coordinates $x,t$ (or more commonly the lightcone 
coordinates $z_\pm$) have been supersymmetrized. However
one can easily realize that such an equation cannot
be derived from a $2D$ manifestly supersymmetric action
principle \cite{IT}. It is neverthless possible to make use 
of a superaction principle where only one coordinate has 
been supersymmetrized, while the other has
been kept ordinary. Such procedure has been employed in
\cite{PST} to define the super-Siegel model \cite{Sie} 
in light-cone coordinates (one made supersymmetric). Here
we adopt the point of view of leaving the time variable
$t$ unchanged while supersymmetrizing the space coordinate
$x$ (now denoted as $X\equiv x, \theta$, with $\theta$ a
Grassmann variable). This approach is quite natural when
dealing with hamiltonian systems which single out the
time coordinate and more suitable for application to 
supersymmetric dual models; indeed one can obtain them from a 
single dynamics instead of being obliged to introduce two 
separate dynamics, one for each chirality.\par
In our conventions $X\equiv x,\theta$ and 
the $N=1$ supersymmetric derivative
is
\begin{eqnarray}
D&=& {\partial\over\partial \theta} +i\theta\partial_1\, , \quad\quad 
\quad D^2 = i\partial_1\, .
\end{eqnarray}
We introduce the superfield $\Phi (X,t)$:
\begin{eqnarray}
\Phi(X,t) &=& \phi(x) + \theta \psi(x)
\, .
\end{eqnarray}
The $N=1$ supersymmetric action $S$ is given by
\begin{eqnarray}
S&=& -i\int dXdt (\partial_0\Phi-\partial_1\Phi ) D\Phi
\, ,
\label{supera}
\end{eqnarray}
which implies for $\Phi$ the equation of motion, equivalent to (\ref{eqmo}),
\begin{eqnarray}
\partial_- D \Phi &=& 0\, .
\end{eqnarray}
In components the supersymmetric lagrangian ${\cal L}_{SFJ}$ is
\begin{eqnarray}
{\cal L}_{SFJ} &=& \partial_0\phi\partial_1\phi - 
(\partial_1\phi)^2 +i \psi
(\partial_0\psi-\partial_1\psi )
\end{eqnarray}
and the global supersymmetry transformation is given by
\begin{eqnarray}
&&\delta\phi = \epsilon \psi\, , \quad\quad
\delta\psi = i \epsilon \partial_1\phi
\, .
\end{eqnarray}
The hamiltonian analysis of the above action can 
be straightforwardly done. At first
we compute the supermomentum $\Pi$
\begin{eqnarray}
\label{103}
\Pi +iD\Phi&=&\Omega \approx  0 \, ,
\end{eqnarray}
which represents one primary superconstraint ($\Omega $)
of second class.\par
The total hamiltonian is
\begin{eqnarray}
\label{104}
H_T &=& H_c+\mu  \Omega \, ,
\end{eqnarray}
 where $\mu$ is an arbitrary multiplier and
\begin{eqnarray}
\label{105}
H_c &=& -i\int dX \partial _1\Phi  D\Phi\, .
\end{eqnarray}
The following generalized Poisson algebra
is satisfied by the superfields
\begin{eqnarray}
\label{106}
\{\Phi (X), \Pi (Y)\}&=& \delta (X,Y)\equiv
\delta (x-y) (\theta_x -\theta_y )
\end{eqnarray}
(all the other Poisson brackets being zero).\par
The constraints (\ref{103}) are found to be second class
\begin{eqnarray}
\label{107} 
&&\Delta (X,Y) =_{def}\{\Omega (X), \Omega (Y)\}= 2iD_X \delta (X,Y)
\label{delta}
\end{eqnarray}
and they do not generate secondary constraints.\par
In order to find the correct equations of motion we 
construct the reduced phase space structure following 
Dirac \cite{Dir}. The inverse of (\ref{delta}) is
\begin{eqnarray}
\Delta^{-1} (X,Y) &=& -{1\over 2} D_X{\partial^{-1}}_x
\delta (X,Y)
\, .
\end{eqnarray}
The Dirac brackets can now be computed for 
any couple of superfields $A(X)$, $B(X)$:
\begin{eqnarray}
\{ A(X), B(Y)\}_D &=& \{A(X), B(Y)\} - \int dZdW
\{A(X), \Omega (Z)\} \Delta^{-1} (Z,W)\{\Omega(W), B(Y)\}
\, .
\nonumber\\
&&
\end{eqnarray}  
In particular we obtain, as fundamental
algebraic relation  
\begin{eqnarray}
\label{108}
\{\Phi (X),\Phi (Y)\}_D&=& {1\over 2} D_X\partial ^{-1}_x
\delta (X,Y)\, .
\end{eqnarray}  
Finally we get the classical hamiltonian equations of motion 
\begin{eqnarray}
\label{109}
&&\partial _0\Phi  (X)=\{\Phi (X),H\}_D= \partial _1\Phi (X)
\, .
\end{eqnarray}
\par 
At the quantum level the equal-time anti-commutator in 
superfield notation is
\begin{eqnarray}
\relax \{ D_X\Phi (X), D_Y\Phi (Y) \}_t &=& -
{1\over 2} D_Y \delta (X,Y)\, ,
\end{eqnarray}
which in components reads 
\begin{eqnarray}
&&\relax [\partial_x \phi(x), \partial_y\phi (y)] 
= -{i\over 2} \partial_y\delta(x-y)\, , 
\quad\quad
\{ \psi (x), \psi (y) \} = {1\over 2} \delta (x-y)\, .
\end{eqnarray}
\par
The supercurrent
\begin{eqnarray}
\vartheta^{00} (X) &=& i : \partial_1\Phi D\Phi : \quad = 
q(x) + \theta l(x)=\nonumber\\
 &=& i:(\partial_1\phi)\cdot \psi: +\theta (-:(\partial_1\phi)^2 :
+i :\partial_1\psi\cdot\psi :)
\end{eqnarray}
gives the $c= {3\over 2}$ $N=1$ superVirasoro algebra. \par
The N\"other conserved charges which generates a constrained
($P^0=P^1$) version
of the $2D$ superPoincar\'e algebra are 
\begin{eqnarray}
&& P=P^0=P^1= \int dx l(x)\, ,\quad\quad M = \int dx (x+t) l(x)\, ,\quad\quad 
Q = \int dx q(x)\, .
\label{superham}
\end{eqnarray}
The non-vanishing (anti)-commutators are
\begin{eqnarray}
&&\relax [ M, P] = i P\, ,\quad\quad  [ M, Q] = i{1\over 2} Q\,, \quad\quad
\{ Q,Q \} = {1\over 2} P\, .
\end{eqnarray}
Just like its bosonic counterpart, 
the supersymmetric action (\ref{supera}) 
is classically invariant under a class of 
$\lambda$-parametrized $1D$-superdiffeomorphisms transformations:
\begin{eqnarray}
\delta \Phi (X,t) &=& \epsilon (z_+,\theta) \partial_1\Phi
-{i\over 2} D\epsilon (z_+,\theta)\cdot D\Phi - {1\over 2} \lambda
\partial_1\epsilon (z_+,\theta) 
\, ,
\label{sdiff}
\end{eqnarray}
where the infinitesimal variation $\epsilon$ is function
of $z_+$, $\theta$ only ($\partial_-\epsilon = 0$).
By performing the same analysis as in the bosonic case
we find the fermionic 
supercurrent $\vartheta_\lambda (X)$ which 
generates the transformations above (\ref{sdiff}):
\begin{eqnarray}
\vartheta_\lambda (X) &=& i : \partial_1\Phi D\Phi : -i \lambda 
\partial_1 D\Phi
\, .
\end{eqnarray}
The (anti)-commutation 
relations satisfied by $\vartheta_\lambda$ produce 
the $N=1$ superVirasoro algebra 
with central extension $ c= {3\over 2 } - 6 i \lambda^2$:
\begin{eqnarray}
\{ \vartheta_\lambda (X), \vartheta_\lambda (Y)\} &=& 
-{1\over 8} (i + 4 \lambda^2) D_y(\partial_y )^2 \delta 
(X,Y)-{3\over 2} i \vartheta_\lambda (Y) \partial_y \delta (X,Y)-\nonumber\\
&&
-{1\over 2} D\vartheta_\lambda (Y)\cdot D_Y \delta (X,Y) -i \partial_y
\vartheta_\lambda (Y)\cdot \delta (X,Y)
\, .
\end{eqnarray}
The conserved charges are computed as in the bosonic case
and the non-anomalous $1D$-superdiffeomorphisms invariance
is recovered for the value $\lambda = \sqrt{ -i\over 4}$.\par
We point out that, for 
the real-valued component fields $\phi$, $\psi$, the equations 
of motion (\ref{eqmo}) can also be obtained from the hamiltonian
\begin{eqnarray}
H&=& -\int dx \left( :(\partial_1\phi)^2 : + i : 
\partial_1\psi \cdot \psi :\right)
\label{hambis}
\end{eqnarray}
with (anti)-commutation relations 
\begin{eqnarray}
&&\relax [\partial_x \phi(x), \partial_y\phi (y)] 
= -{i\over 2} \partial_y\delta(x-y)\, , \quad\quad
\{ \psi (x), \psi (y) \} = -{1\over 2} \delta (x-y)\, .
\end{eqnarray}
The above formulas are recovered from the previous ones 
after setting $\psi \mapsto i \psi$.\par
The hamiltonian $H$ of eq. (\ref{hambis}) however, unlike 
$P^0$ in eq. (\ref{superham}), is not 
supersymmetric because the fermionic 
hermitian operator $Q= i \sqrt{2} :\partial_1\phi\psi :$ in this case
leads to $\{Q,Q\}  = - H$. Notice the presence of the
``wrong" minus sign. When dealing with extended 
supersymmetries or supersymmetric dual models one has to be very careful
in picking up the ``correct" supersymmetric hamiltonian.\par
We devote the last part of this section to discuss the
superextension of the Tseytlin model. As in the bosonic
case the supersymmetric dual model can be decomposed
into two independent, respectively chiral and 
antichiral, $N=1$ FJ models. The supersymmetric 
action which coincides with a dimensional reduction of  
the $4$-dimensional super-Schwarz-Sen model is up to a normalizing factor
\cite{CD} 
\begin{eqnarray}
\label{100}
S = \int d^2x \left[ \partial_{0} \phi \partial_{1} 
{\tilde \phi}+
\partial_{0} {\tilde\phi } \partial_{1} \phi  - 
(\partial_{1} \phi)^2  - (\partial_{1} {\tilde\phi})^2     
\right.  \nonumber\\
 \left. + i\psi \partial _0\psi 
 + i{\tilde \psi}\partial _0{\tilde \psi} -i\psi 
\partial _1 {\tilde\psi}
-i{\tilde\psi} 
\partial _1 \psi 
\right]\,\,\, , 
\end{eqnarray}
and can be rewritten in superfield notations as
\begin{eqnarray}
S &=& -i \int dX dt \left( \partial_0 \Phi_+
-\partial_1\Phi_+ )D\Phi_+ -(\partial_0 \Phi_- 
+\partial_1\Phi_- )D\Phi_-\right)
\, ,
\end{eqnarray}
where
\begin{eqnarray}
&&\Phi_+ = \phi_+ + \theta \psi_+\, , \quad\quad
\Phi_- = \phi_- + i\theta \psi_-\, 
\end{eqnarray}
are chiral (antichiral) superfields and 
\begin{eqnarray}
&& \phi_\pm = {1\over \sqrt{2}} (\phi\pm 
{\tilde \phi})\, ,\quad\quad
\psi_\pm = {1\over \sqrt{2}} (\psi\pm {\tilde\psi}) 
\, .
\end{eqnarray}
Notice that the presence of an extra ``$i$" in 
the decomposition of $\Phi_-$ is in order to make
the hamiltonian for the antichiral sector supersymmetric,
as explained above.\par
The duality invariance corresponds to the exchange 
$\Phi_\pm \leftrightarrow  \pm \Phi_\pm $.\par
The N\"other analysis is recovered from the results 
of the $N=1$ FJ model. The anticommutation relations
are
\begin{eqnarray}
\relax [ D\Phi_\pm (X), D\Phi_\pm (Y) ] &=& \mp {1\over 2}
D_Y\delta (X,Y) , \quad\quad [ D\Phi_+ (X), D\Phi_- (Y) ] =0\, .
\end{eqnarray}
Two independent $c={3\over 2} $ superVirasoro algebras
result from the supercurrents $\vartheta_\pm $:
\begin{eqnarray}
&& \vartheta_\pm = \gamma_\pm :\partial_1\Phi_\pm\cdot D\Phi_\pm :\quad = 
q_\pm -i \gamma_\pm \theta l_\pm
\, ,
\end{eqnarray}
where $\gamma_+ = i$ and $\gamma_- = -1$.\par
Let us introduce the currents
\begin{eqnarray}
&&\vartheta^{00} = l_++l_-, \quad\quad 
\vartheta^{01} = l_+-l_-,\quad\quad q^{01} = q_++q_-,
\quad\quad q^{02} = q_+-q_-\, .
\end{eqnarray}
The superPoincar\'e algebra is realized by the bosonic
conserved charges
\begin{eqnarray}
&& P^0 = \int dx \vartheta^{00},\quad\quad P^1 =\int 
dx \vartheta^{01},\quad\quad
M = \int dx (x \vartheta^{00} + t \vartheta^{01})
\, ,
\end{eqnarray}
together with the supercharges 
\begin{eqnarray}
Q^1 =\int dx q^{01}, \quad\quad Q^2 =\int dx q^{02}
\, .
\end{eqnarray}    
As a quantum mechanical system, the 
``$N=1$" Tseytlin model is globally $N=2$ supersymmetric 
since $Q^{1,2}$ satisfy
\begin{eqnarray}
\{Q^1,Q^1\}=\{Q^2,Q^2\}= H,\quad\quad \{Q^1,Q^2\} = 0
\, ,
\end{eqnarray}
where $H$ is the hamiltonian.\par
Explicitly $Q^1$, $Q^2$ generate two 
supersymmetry transformations \begin{eqnarray}
&&
\delta_1\phi_\pm = {\epsilon_1\over 2} \psi_\pm, \quad\quad \delta_1\psi_\pm 
= \pm 
{i\over 2}\epsilon_1\partial_1\phi_\pm ; \quad\quad
\delta_2\phi_\pm =\pm {\epsilon_2\over 2} \psi_\pm,
\quad\quad
\delta_2\psi_\pm = {i\over 2}\epsilon_2\partial_1\phi_\pm
\, .
\end{eqnarray}
The model is superconformally invariant and 
non-anomalous even in the quantum case as a 
trivial consequence of the $1D$ superdiffeomorphisms 
invariance of the $N=1$ FJ theory.  
\vspace{0.2cm}
\noindent{\section{The $N=2$ extensions.}} 
\par
The $N=2$ extensions of the FJ and the (globally $N=4$ invariant) 
Tseytlin model can be constructed by 
mimicking the previous constructions in a 
manifest $N=2$ superfield formulation. Since the analysis proceeds
as before we limit ourselves to write the results.\\
When dealing with $N=2$ superfields we have at first to
establish if real or constrained (anti)-chiral
superfields are employed. It turns out 
that chiral-antichiral superfields make the job.\par
The theories will be defined by leaving the time $t$ 
ordinary while the space coordinate will be $N=2$
supersymmetrized with the introduction of $\theta,
{\overline\theta}$ Grassmann variables. Our $N=2$
conventions (see also \cite{IT}) are as follows. The
fermionic derivatives $D, {\overline D}$ are 
\begin{eqnarray}
&& D= {\partial\over\partial\theta}-{i\over 2}
{\overline\theta}\partial_1\, , \quad\quad
{\overline D} = {\partial\over\partial{\overline\theta}}
-{i\over 2} \theta\partial_1\, .
\end{eqnarray}
They satisfy the equations
\begin{eqnarray}
&& D^2={\overline D}^2 = i\partial_1\, , \quad\quad
\{ D, {\overline D}\} = 0\, .
\end{eqnarray}
$N=2$ chiral ($\Phi$) and antichiral
(${\overline\Phi}$) superfields satisfy
the constraints
\begin{eqnarray}
&& D\Phi=0 \, ,\quad\quad {\overline D}{\overline\Phi}=0
\, .
\end{eqnarray}
In components we have
\begin{eqnarray}
&&\Phi = \phi +{\overline\theta} \psi + {i\over 2}
\theta{\overline\theta}\partial_1\phi\, , \quad\quad
{\overline\Phi}= {\overline\phi} + \theta{\overline\psi}
-{i\over 2} \theta{\overline\theta} \partial_1{\overline\phi}\, .
\end{eqnarray}
The $N=2$-invariant action for the FJ model is
given by the sum of two pieces involving 
separately $N=2$ chiral and antichiral superfields:
\begin{eqnarray}
S&=& i \int dt dX_L \left(\partial_0\Phi -\partial_1\Phi
) D{\overline\Phi}\right) + i \int dt dX_R \left(
(\partial_0{\overline\Phi}-\partial_1{\overline\Phi})
{\overline D}\Phi\right)
\, ,
\end{eqnarray}
where $dX_L\equiv dxd\theta$, $dX_R\equiv dx d{\overline
\theta}$ denote integration over the chiral (antichiral)  
variables.\par
In components the lagrangian ${\cal L}$ is
\begin{eqnarray}
{\cal L} &=& \partial_0\phi \partial_1{\overline\phi}
+\partial_0{\overline\phi}\partial_1\phi - 2\partial_1\phi
\partial_1{\overline\phi} +
i(\partial_0\psi-\partial_1\psi){\overline \psi}
+ i(\partial_0{\overline\psi} -\partial_1{\overline\psi})
\psi
\, .
\end{eqnarray}
The reality condition sets ${\phi}^\dagger = {\overline
\phi}$, ${\psi}^\dagger = {\overline\psi}$.\par
At the level of the equations of motion we obtain two
copies of the supersimmetric FJ equations. Indeed
\begin{eqnarray}
&&\partial_1\partial_-\phi =\partial_1\partial_-{\overline\phi} 
=0\, , \quad\quad\partial_-\psi=\partial_-{\overline\psi} =0 \, .
\end{eqnarray}
The (anti)-commutation relations which define the
hamiltonian dynamics are given by
\begin{eqnarray}
&&\relax [ \partial_1{\overline \phi}(x), \partial_1\phi
(y) ] = {i\over 2} \partial_y\delta (x-y)\, , \quad\quad
\{{\overline{\psi}} (x), \psi (y) \} = {1\over 2} \delta
(x-y)\, .
\end{eqnarray}
and vanishing otherwise.\par 
In manifest $N=2$ superfield notation they are written as
\begin{eqnarray}
\{ D{\overline\Phi} (X), {\overline D}\Phi(Y) \}
&=& {1\over 2} D_X {\overline D}_Y \delta (X,Y)
\, ,
\end{eqnarray}
here $\delta (X,Y) = \delta (x-y) (\theta_x-\theta_y)
({\overline\theta}_x -{\overline\theta}_y) $ is the
$N=2$ supersymmetric delta-function.\par
The hamiltonian $H$ is given by
\begin{eqnarray}
H &=& \int dt dX  : D{\overline \Phi}\cdot{\overline D}\Phi:
\quad .
\end{eqnarray}
The supercurrent $ J(X) =   : D{\overline \Phi}\cdot{\overline D}\Phi: =
j + \theta q + {\overline \theta} {\overline q} 
+ \theta{\overline\theta} l $,
where
\begin{eqnarray}
&&j(x)= :\psi{\overline\psi}:\, , \quad\quad
q(x) = i:\partial_1\phi\cdot
{\overline\psi} : \, , \quad\quad 
{\overline q} (x) = -i:  \partial_1{\overline\phi}
\cdot\psi :\, ,\nonumber\\
&&l(x) = :\partial_1\phi\cdot\partial_1{\overline\phi}: +
{i\over 2}:{\partial_1\psi\cdot{\overline\psi}}:+{i\over 2}
:{\partial_1{\overline\psi}}\cdot\psi :
\end{eqnarray}
satisfy the $N=2$ superVirasoro algebra with central 
charge $c=3$ in the quantum case (the standard 
OPE conventions are recovered by rescaling 
$l\mapsto {\tilde l} = - 2 i l,\quad 
(q,{\overline q}) \mapsto ({\tilde q}, {\tilde{\overline q}}) =
 \sqrt{-8 i} (q, {\overline q}), \quad
j \mapsto {\tilde j} = 2 j$):
\begin{eqnarray}
\relax [ l(x), l(y) ] &=& -{3\over 48}{\partial_y}^3\delta(x-y) + i 
l(y)\partial_y\delta(x-y)
+{i\over 2} \partial_y l(y)\cdot \delta (x-y)\, ,\nonumber\\
\relax [l(x), q(y)] &=& {3 i\over 4} q(y) \partial_y\delta
(x-y) +{i\over 2} \partial_y q(y)\cdot \delta (x-y)\, ,\nonumber\\
\relax[l(x), {\overline q}(y)] &=& {3 i\over 4} 
{\overline q}(y) \partial_y\delta(x-y)+{i\over 2} 
\partial_y {\overline q}\cdot (y)\delta(x-y)\, , \nonumber\\
\relax [ l(x), j(y) ] &=& {i\over 2} j(y) \partial_y\delta
(x-y) +{i\over 2}\partial_yj(y)\cdot\delta (x-y)\, ,\nonumber\\
\{ q(x), {\overline q}(y)\} &=& {i\over 8} {\partial_y}^2
\delta (x-y) +{i\over 2} j(y)\partial_y\delta (x-y) +
({1\over 2} l(y) +{i\over 4} \partial_y j(y))\cdot \delta (x-y)\, ,\nonumber\\
\relax [j(x), q(y) ] &=& {1\over 2} q(y) \delta (x-y)\, , 
\quad\quad [ j(x) , 
{\overline q}(y)] = -{1\over 2} {\overline q}(y)\delta (x-y)\, ,\nonumber\\
\relax[j(x), j(y) ] &=& {1\over 4} \partial_y\delta (x-y)
\end{eqnarray}
and vanishing otherwise.\par
The above currents are the building blocks to 
construct the $N=2$ superPoincar\'e 
generators just like in the previous cases. \par
In particular the $N=2$ global hermitian charges are 
\begin{eqnarray}
Q_1 =_{def} \int dx \left( q(x) +{\overline q} (x)\right)
\, ,
\quad && \quad Q_2 =_{def} i \int dx \left( q(x) - 
{\overline q}(x) \right)\, ,\nonumber
\end{eqnarray}
which satisfy $\{Q_1, Q_2\} = 0$, $ \{ Q_1,Q_1\} =\{Q_2,Q_2\} = H$.\par 
The (non-anomalous) invariance under $1$-dimensional
$N=2$ diffeomorphisms is 
implied in the quantum case by the existence of modified currents
$u_{\lambda,{\overline\lambda}}$ ($u$ denotes either $j$, $q$, 
${\overline q}$ or $l$),
which satisfy an $N=2$ superVirasoro where all central
charges are vanishing. 
The modified currents cannot be accomodated into an
$N=2$ superfield formalism and we are obliged to write 
them in components. We get
\begin{eqnarray}
j_{\lambda ,{\overline\lambda}}(x)&=&:\psi{\overline\psi}:
-i\lambda{\partial_1\phi} -i{\overline\lambda}\partial_1{\overline\phi}\, ,
\nonumber\\
 q_{\lambda ,{\overline\lambda}}(x) &=& i:\partial_1\phi\cdot
{\overline \psi} : - i{\overline \lambda} \partial_1 
{\overline \psi}\, , \quad\quad
{\overline q}_{\lambda,{\overline\lambda}} (x) = -i:  
\partial_1{\overline\phi}
\cdot\psi :    - i {\lambda} \partial_1\psi
\, ,\nonumber\\
l_{\lambda,{\overline\lambda}}(x) &=& :\partial_1\phi\cdot\partial_1
{\overline\phi}: +
{i\over 2}:{\partial_1\psi\cdot{\overline\psi}}:+{i\over 2}
:{\partial_1{\overline\psi}}\cdot\psi :
+{\lambda\over 2}{\partial_1}^2 \phi -{{\overline\lambda}\over 2}
{\partial_1}^2 {\overline\phi}
\, .
\end{eqnarray}
If $\lambda$, ${\overline\lambda}$ are chosen in such a way 
that $\lambda\cdot {\overline\lambda} = -{i\over 4}$
then all central charges are vanishing; as an example 
in particular
\begin{eqnarray}
\relax [j_{\lambda,{\overline\lambda}} 
(x), j_{\lambda,{\overline\lambda}} (y) ] &=& -i 
(\lambda{\overline\lambda} + {i\over 4})\partial_y\delta (x-y)
\, .
\end{eqnarray}
We notice that the modified current $l_{\lambda,{\overline\lambda}}$ is 
no longer 
hermitian if the above constraint is taken into 
account (however on abstract level the closed algebraic 
structure it satisfies is compatible 
with a hermitian condition). This feature is not specific 
of $N=2$ but is already present in the bosonic and $N=1$ cases.\par
The modified currents $u_{\lambda,{\overline\lambda}}$
are the generators of the $N=2$
$1$-dimensional diffeomorphisms 
invariances, the infinitesimal transformation being 
given by the commutators with
\begin{eqnarray}
&&\int dx \epsilon_u (z_+) u_{\lambda,{\overline\lambda}} (x)\, .
\end{eqnarray}
The dualized version of the $N=2$ FJ model is now easily
constructed by introducing a second set of superfields
(antichiral in spacetime). The action is given by 
\begin{eqnarray}
S&=& 2 i \int dt dX_L \left(\partial_-\Phi_+ 
\cdot D{{\overline\Phi}_+} -\partial_+\Phi_-\cdot 
D{\overline \Phi}_-\right) + 2 i \int dt dX_R \left(
\partial_-{{\overline\Phi}_+}\cdot
{\overline D}\Phi_+ -\partial_+{\overline\Phi}_-\cdot 
{\overline D}\Phi_- \right)\, .\nonumber\\
&&
\end{eqnarray}
The correct expansion for $\Phi_\pm$, ${\overline\Phi}_\pm$ in 
hermitian component fields
which leads to the supersymmetric hamiltonian (see the remark
in section $3$) is given by:
\begin{eqnarray}
&& 
\Phi_\pm = \phi_\pm -i \gamma_\pm {\overline\theta}
\psi_\pm + {i\over 2}\theta{\overline\theta}  \partial_1\phi_\pm\, , 
\quad\quad
{\overline\Phi}_{\pm} ={\overline\phi}_\pm -i\gamma_\pm
\theta {\overline\psi}_\pm -{i\over 2}\theta{\overline\theta} 
\partial_1{\overline \phi}_\pm
\end{eqnarray}
(here again $\gamma_+ = i$, $\gamma_- = -1$).\par
The lagrangian in components reads
\begin{eqnarray}
{\cal L} &=& 2( \partial_-\phi_+\cdot\partial_1{\overline\phi}_+ +
\partial_-{\overline\phi}_+ \cdot\partial_1\phi_+ 
-\partial_+\phi_-\cdot \partial_1{\overline\phi}_-
-\partial_+{\overline\phi}_-\cdot\partial_1\phi_-+
\nonumber\\
&& + i \partial_-\psi_+\cdot{\overline\psi}_+ + i
\partial_-{\overline\psi}_+\cdot\psi_+
+i\partial_+\psi_-\cdot{\overline\psi}_- +
i\partial_+{\overline\psi}_-\cdot\psi_-)
\, .
\end{eqnarray}
The non-vanishing (anti)-commutators are 
\begin{eqnarray}
&&\relax [ \partial_x{\overline\phi}_\pm (x), \partial_y
{\phi}_{\pm}(y) ] = \pm {i\over 2} \partial_y\delta (x-y)
\, , \quad\quad
\{ {\overline\psi}_\pm (x), \psi_\pm(y) \} = {1\over 2}
\delta (x-y)
\, .
\end{eqnarray}
The conserved currents of the antichiral sector generate a
second $N=2$ superVirasoro algebra with (quantum) central charge $c=3$.\par
Following the same reasoning as in 
the previous  section we can construct 
four global supercharges $Q^i$ ($i=1,...,4$) 
which lead to a global $N=4$ supersymmetry ($\{Q^i, Q^i\}= H$ for any $i$ and
$\{Q^i, Q^j \} =0$ for $i\neq j$). The Coulomb gas 
realization for the $N=2$ FJ model implies
the full $N=2$ superconformal invariance for the quantum dual model. 

\vspace{0.2cm}
\noindent{\section{Conclusions.}}

\par
In this paper $N=1,2$ extensions of chiral and dual 
models have been constructed and their symmetry properties
analyzed. In particular their relativistic
character was proven by computing global charges 
which close the $2D$-superPoincar\'e algebra. 
The invariances under $1D$-superdiffeomorphisms and 
respectively superconformal transformations were 
furthermore investigated. It was shown that, due to $N=1,2$ 
Coulomb gas results, modified currents exist 
which lead to non-anomalous quantum theories.\par
The present work was mainly motivated to
establish an algebraic framework for dimensional reductions of 
higher-dimensional
supersymmetric dual models. The algebraic structures found
however have an interest in their own and further
investigations look promising. 
Currently under study, e.g. the $N=4$ supersymmetric 
extensions seem related to non-abelian 
structures leading to a new $N=4$ realization of the Coulomb gas.\par
Another topic which deserves to be studied, as 
suggested in \cite{Sen}, is the coupling of the 
above theories with $2D$ supergravity, with special attention to the
presence of anomalies.

\end{document}